\documentclass[aps,prb,reprint,nofootinbib,twocolumn,superscriptaddress,showpacs,showkeys,longbibliography,amsmath,amssymb]{revtex4-2}
\usepackage{graphicx}
\usepackage{dcolumn}
\usepackage{bm}
\usepackage{braket}
\usepackage{subfigure}
\usepackage[colorlinks,bookmarks=false,citecolor=blue,linkcolor=red,urlcolor=blue]{hyperref}
\usepackage[english]{babel}
\usepackage{times}
\usepackage{helvet}
\usepackage{courier}

\makeatletter

\newcommand{\Rmnum}[1]{\expandafter\@slowromancap\romannumeral #1@}
\makeatother

\begin{document}
\preprint{APS/123-QED}

\title{Topological laser in a two-dimensional Su-Schrieffer-Heeger lattice with artificial gauge flux } 

\author{Yi-Ling Zhang}
\affiliation{School of Physical Science and Technology, and Collaborative Innovation Center of Suzhou Nano Science and Technology, Soochow University, 1 Shizi Street, Suzhou 215006, China}
\author{Yang Liu}
\affiliation{School of Physical Science and Technology, and Collaborative Innovation Center of Suzhou Nano Science and Technology, Soochow University, 1 Shizi Street, Suzhou 215006, China}
\author{Zhi-Kang Lin}
\email{zklin@hku.hk}
\affiliation{Department of Physics, The University of Hong Kong, Hong Kong 999077, China}
\author{Jian-Hua Jiang}
\email{jhjiang3@ustc.edu.cn}
\affiliation{School of Physical Science and Technology, and Collaborative Innovation Center of Suzhou Nano Science and Technology, Soochow University, 1 Shizi Street, Suzhou 215006, China}
\affiliation{School of Biomedical Engineering, Division of Life Sciences and Medicine, University of Science and Technology of China, Hefei 230026, China}
\affiliation{Suzhou Institute for Advanced Research, University of Science and Technology of China, Suzhou, 215123, China}

\date{\today}

\begin{abstract}
Topological lasers, known for their robustness and unique features originating from nontrivial topology, have recently become a focal point of research in photonics. In this work, we propose a topological laser based on two-dimensional Su-Schrieffer-Heeger photonic lattices as induced by artificial gauge flux insertion. The underlying effect, called the topological Wannier cycles, is characterized by topological local modes with continuously tunable frequency and orbital angular momentum emerging in two photonic band gaps. These topological local modes enable single-mode large-area lasing in each photonic band gap with both topological robustness and exceptional tunability in frequency and OAM properties, setting a notable contrast with previous topological lasers. We further discuss both localized and extended artificial gauge flux insertion and compare their properties. We find that extended gauge flux achieves significantly higher laser output intensity and larger single-mode area under laser-gain conditions, outperforming the local gauge flux configuration in both output intensity and resilience against disorders. We also elucidate the precise mechanisms by which nonlinear gain and gauge flux govern the photon dynamics in various regimes. These results provide crucial theoretical insights for OAM control in topological lasers and pave the way for advancements in high precision engineering of lasers and optical systems.

\end{abstract}

\maketitle

\section{Introductioon}
With the rapid development of topological photonics, topological lasers have emerged as cutting-edge technology and have garnered significant attention in both photonic and condensed matter communities. Topological photonics focused mainly on the realization of topologically protected boundary states, such as topological edge states with time-reversal symmetry breaking ~\cite{topological_Photonic1, topological_Photonic2, Hafezi2011, Hafezi2013, topological_Photonic3, PhysRevB.98.205147, PhysRevLett.122.233903, topological_Photonic4, topological_Photonic5, Xie2021}, topological bound states induced by topological defects, i.e., Dirac vortexes, disclinations and dislocations that inherently represent structural distortions~\cite{topological_disclination1, topological_disclination2, topological_disclination3, Lin2023}, and unidirectional light propagation allowed in Floquet-driving or subtly designed waveguides~\cite{topological_waveguides1, topological_waveguides2, topological_waveguides3, topological_waveguides4}. These defect immune topological photonic states are vital for optical computing and communication and may facilitate the efficiency for photonic device. Building on this concept, research on topological lasers has demonstrated single-mode lasing through topological edge states. Meanwhile, theoretical proposals and experimental realizations of topological laser emission have been achieved in various topological systems~\cite{topological_laser1, topological_laser2, topological_laser3, topological_laser4, topological_laser5}, as well as Kekule-modulated lattices that give rise to zero- dimensional vortex defect states. These studies have not only revealed the potential applications of topological edge states under optical gain conditions but also offered new approaches to addressing technological challenges, such as large area, high power coherent emission.

Recently, topological Wannier cycles, characterized by cyclic defect localized spectral flows traversing topological band gaps, have provided new insights into topological phenomena in crystalline insulators~\cite{topological_Wannier_cycles1}. These cycles, which emerge in systems with both local gauge flux and nontrivial real space topological invariants (RSTIs), have been extensively explored in various synthetic materials, such as photonic and acoustic crystals, where local gauge flux insertion has been experimentally realized in single unit-cell plaquettes~\cite{Lin2023, Lin2022, topological_Wannier_cycles2, topological_Wannier_cycles3, Kong_2022}. Notably, the ability to dynamically manipulate Wannier cycles via artificial gauge fields has not only advanced our understanding of topological matter but also introduced a promising avenue for topological laser design. The incorporation of topological Wannier cycles offers distinct advantages and more degrees of freedom for designing topological lasers, unlike traditional topological lasers that rely primarily on topological chiral edge states~\cite{topological_laser3, topological_laser4, PhysRevResearch.4.013195, Ota2018, Zhang2020, Tian2023, doi:10.1126/sciadv.adg4322}, valley edge modes~\cite{PhysRevResearch.1.033148, PhysRevB.103.245305, doi:10.1126/science.aao4551}, and topological Dirac-vortex states~\cite{Yang2022, Ma2023, Liu2024}. To this end, we propose, for the first time, a topological laser design based on topological Wannier cycles, which truly enables highly tunable lasing properties.

Specifically, the topological laser design is demonstrated in a two-dimensional (2D) Su-Schrieffer-Heeger (SSH) lattice, utilizing topological Wannier cycles with local gauge flux (LGF) and additional gain modulation. Moreover, the original local gauge flux is generalized to encompass a larger spatial volume, thereby establishing a closer physical analogy to the gauge flux pattern induced by screw dislocations in natural solid materials. The extended gauge flux (EGF) configuration further comforms and exploits the cyclic evolution of eigenstates across topological band gaps induced by local gauge flux. The detailed analysis reveals that the resultant laser is of single mode with a center localized profile, which not only retains the robustness inherent to topological systems, but also demonstrates highly control of frequency and OAM modulation. Our design achieves precise continuous tunability over OAM carrying laser modes through localized gain and flux distributions, thereby enabling the generation of high power coherent emission with tunable OAM. This property is particularly advantageous for quantum information processing, high-dimensional encoding, and advanced optical communication~\cite{oam1, oam2, oam3, oam4, oam5, oam6}.

Our main results are as follows: In Sec. \ref{sec: the model}, we review the 2D SSH lattice model and introduce the theoretical framework for gain, losses, and artificial gauge flux. 
In Sec. \ref{sec.LOCALIZED GAIN AND GAUGE FLUX (LGF)}, we analyze the scenario where gain and flux are localized in the SSH lattice. We first examine topological Wannier cycles by investigating the spectral flow and cyclic evolution of eigenstates across the band gaps. This is followed by an exploration of single-mode laser emission, focusing on how the laser mode retains OAM information, with fidelity analysis demonstrating that the mode corresponds to the original eigenstate when fidelity reaches 1.
In Sec. \ref{sec:EXTENDED GAIN AND GAUGE FLUX (EGF)}, we shift our focus to a extended gain and flux configuration. Similar to the LGF configurations, we analyze the topological Wannier cycles, explore the conditions for single mode lasing, and study the evolution and selection of OAM carrying laser modes. Even with stronger mode competition, the s-state consistently dominates.
In Sec. \ref{subsec.Mode selection2}, we compare the LGF and EGF configurations, highlighting the EGF’s superior OAM tunability and higher single mode intensity six times greater than the LGF configuration.
Finally, in Sec. \ref{sec:single-mode area}, we discuss the larger single mode area of the EGF configuration, demonstrating its superior efficiency and stability. Conclusions are drawn in Sec. \ref{sec:DISCUSSION AND CONCLUSION}.

\section{MODEL}
\label{sec: the model}
In this work, we introduce two types of artificial gauge fluxes with different spatial scales into a 2D SSH optical lattice model, aiming to reveal and understand novel topological laser phenomena. We begin by reviewing the 2D SSH model, whose Hamiltonian in real space is defined as follows ~\cite{PhysRevLett.118.076803, PhysRevB.100.075437}:
\begin{align}
H_{\text{ssh}}&=\sum_{m,n}[J(1+(-1)^m\delta )a^{\dagger}_{m,n}a_{m+1,n}\notag\\
&+J(1+(-1)^n\delta )a^{\dagger}_{m,n}a_{m,n+1}]+\text{h.c.},
\label{Hamiltonian_total}
\end{align}
where \( a_{m,n} \) and \( a^{\dagger}_{m,n} \) denote the annihilation and creation operators of the photonic field at lattice site \((m,n)\), respectively. The coupling constants \( J_1 = J(1-\delta) \) and \( J_2 = J(1+\delta) \) correspond to intra-cell and inter-cell hopping amplitudes, and the parameter \(\delta\) controls the topological phase of the system.
\begin{figure}[]
	\centering
	\includegraphics[width=0.48\textwidth]{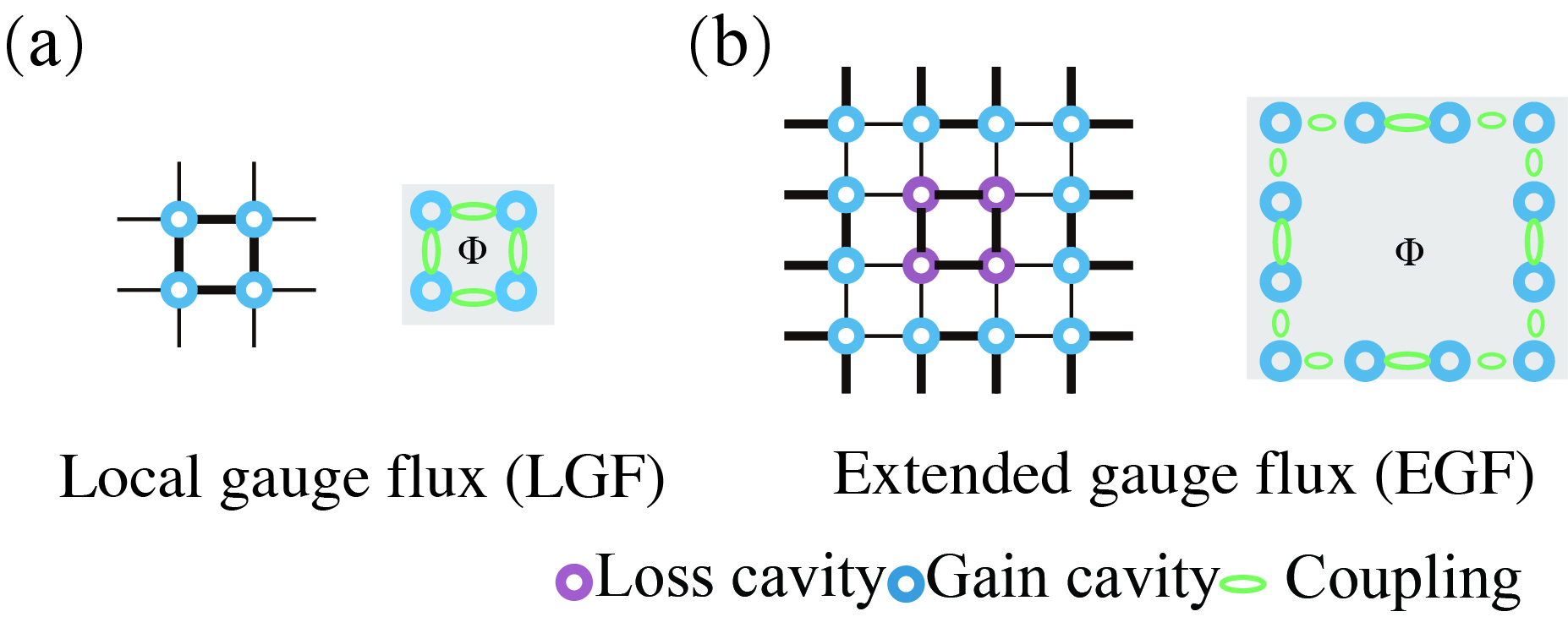}
\caption{(a) Left: LGF structure localized at the center; Right: Schematic diagram of the structure composed of microcavities and coupling cavities as implemented in experiments. (b) Left: EGF structure on a larger spatial scale; Right: Schematic diagram of the structure composed of microcavities and coupling cavities as implemented in experiments.}
\label{fig:flux}
\end{figure}

Notably, in the topological phase of the two-dimensional SSH superlattice, the energy band structure can be naturally described by Wannier orbitals localized in real space. When an artificial gauge flux \(\Phi\) is locally introduced in the central region of the lattice and continuously tuned from \(0\) to \(2\pi\), the system exhibits filling anomalies under various lattice symmetries~\cite{Lin2022}. These filling anomalies manifest as a continuous gapless spectral flow, whose physical origin is the topological reconstruction and cyclic evolution of the Wannier orbitals near the lattice center induced by the localized gauge flux. Such a phenomenon is therefore termed the topological Wannier cycle. The filling anomalies mentioned above can be quantitatively characterized by real-space topological invariants. Intuitively, the topological Wannier cycle can also be understood as the modification of the geometric phase acquired by photons propagating along closed loops due to the artificial gauge flux, analogous to the Aharonov–Bohm effect in electronic systems, thus reflecting robust topological dynamics.

In addition, two artificial gauge flux configurations with different spatial scales and experimental feasibility are proposed, as illustrated in Fig. \ref{fig:flux}. The first configuration, termed the LGF, is illustrated in Fig. \ref{fig:flux}(a). It consists of a \(2\times2\) microcavity array at the lattice center, characterized by strong inter-cell couplings forming a closed loop that generates a localized nonzero flux \(\Phi\). The second configuration, the EGF shown in Fig. \ref{fig:flux}(b), involves a \(4\times4\) microcavity loop with alternating weak intra-cell and strong inter-cell couplings, creating an extended region where photon propagation accumulates a controllable nonzero gauge flux, potentially enabling richer topological defect states. Experimentally, this can be achieved by precisely controlling the coupling phases between microcavities. Such control can be achieved using nonreciprocal optical elements (e.g., Faraday rotators) integrated into coupling paths or through dynamic modulation methods like Floquet engineering, which periodically vary refractive indices or cavity lengths~\cite{10.1063/5.0061950, PRXQuantum.5.040331}.

Finally, based on the two-dimensional SSH lattice model incorporating two distinct types of artificial gauge fluxes, we investigate novel topological laser phenomena by taking into account both non-Hermitian and nonlinear effects. Within the framework of semiclassical laser theory for broadband gain media, the time evolution of the classical photonic field amplitude is governed by a nonlinear dynamical equation that incorporates both gain and loss \cite{topological_laser1}:
\begin{equation}
\dot{a}_{m,n}(t) = -i[a_{m,n}, H_{\text{SSH}}] + \left( \frac{gP_{m,n}}{1 + |a_{m,n}|^2 / I_{\text{sat}}} - \gamma \right) a_{m,n},
\label{laser_dynamics}
\end{equation}
where \(H_{\text{SSH}}\) is defined as in Eq.~(\ref{Hamiltonian_total}), \(gP_{m,n}\) denotes the spatially distributed gain, \(\gamma\) is the intrinsic loss of the resonators, and \(I_{\text{sat}}\) represents the saturation intensity.

\section{LOCALIZED GAIN AND GAUGE FLUX (LGF)}
\label{sec.LOCALIZED GAIN AND GAUGE FLUX (LGF)}
\subsection{Topological Wannier cycle in LGF}
\label{sec.Topological Wannier cycle in LGF}

\begin{figure}[]
	\centering
	\includegraphics[width=0.3\textwidth]{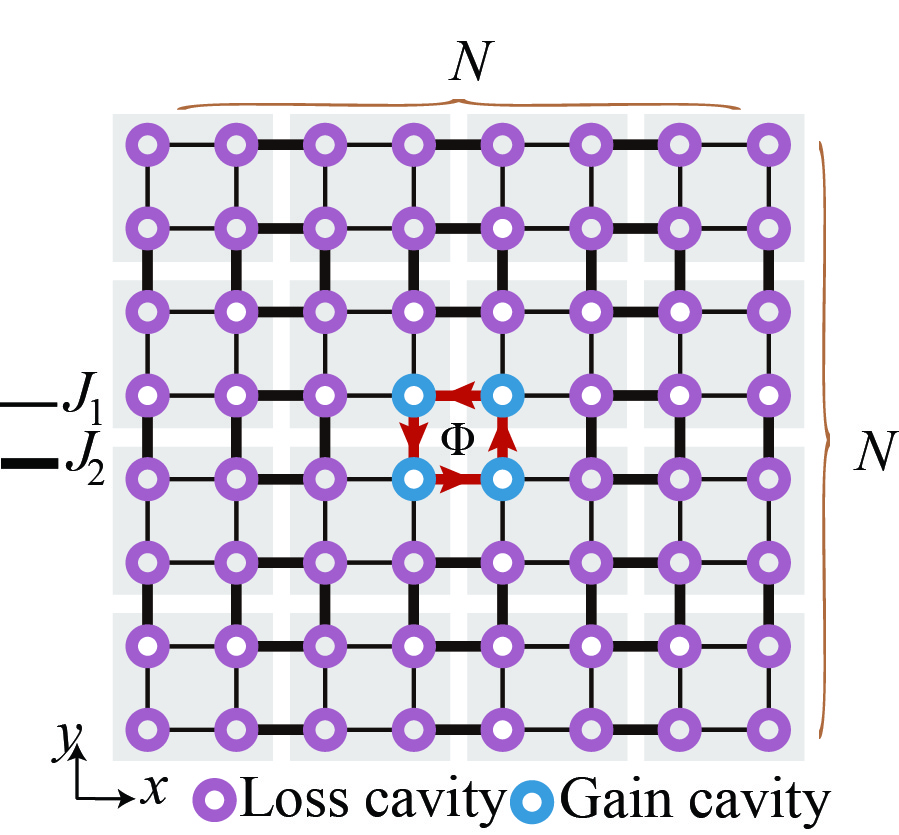}
\caption{A planar Lattice with central gauge flux representation of the SSH model in Eq. (\ref{Hamiltonian_total}) with open boundaries in both directions. The sites has a particle loss (red) and a particle gain (blue). $\Phi$ is an adjustable periodic magnetic flux in the middle of the plate, the $J_1$ and $J_2$ is the intra-cell and inter-cell coupling, respectively. }
\label{fig:center_flux}
\end{figure}

First, we investigate the topological Wannier cycle induced by the introduction of a LGF in the 2D SSH lattice model. As shown in Fig.~\ref{fig:center_flux}, the artificial gauge flux is applied at the center of the lattice, leading to an additional phase factor in the Hamiltonian:
\begin{equation}
\langle m,n,\alpha|H(\Phi)|m',n',\beta \rangle=e^{i\theta/4}\langle m,n,\alpha|H_{\mathrm{ssh}}|m',n',\beta \rangle,
\label{eq:benzhengzhi}
\end{equation}
where \(|m,n,\alpha\rangle\) and \(|m',n',\beta\rangle\) denote states on the \(\alpha\) and \(\beta\) sublattices of the unit cells located at \((m,n)\) and \((m',n')\), respectively. \(H(\Phi)\) and \(H_{\text{ssh}}\) denote the system Hamiltonian with the gauge flux and the original Hamiltonian of the 2D SSH model, respectively. It can be seen that by traversing a closed loop the total magnetic flux \(\Phi\) is obtained, which is analogous to the Aharonov–Bohm effect: the wave function accumulates a phase \(e^{i\theta}\) when encircling the magnetic flux region. In other regions of the lattice, the total gauge flux is zero, as it is gauged away. If the original 2D SSH model is in a topologically nontrivial phase, the introduction of the gauge flux may induce new topological phenomena.
In the absence of the artificial gauge flux, we examine the eigenvalues and corresponding orbital types of the eigenfunctions under the action of the \(\hat{C}_4\) rotation operator. When the symmetry operator \(\hat{C}_4\) acts on the eigenfunction, it satisfies \(\hat{C}_4|\psi\rangle=E_n^4|\psi\rangle\). Since \(C_4^4=I\), the corresponding eigenvalues are given by \(E_n=e^{i\frac{2\pi n}{4}},~n=0,1,2,3(-1)(\text{mod}~4)\). If the eigenfunction is an eigenstate of the orbital angular momentum \(\hat{L}_z\), the eigenvalues are \(1,i,-1,-i\), corresponding to the orbital angular momenta \(l_z^{n}=n=0,1,2,3(-1)(\text{mod}~4)\) and the orbital types of the eigenstates are \(s,\; p_+(p_x+ip_y),\; d,\; p_-(p_x-ip_y)\), respectively.

When an artificial gauge flux \(\Phi\) is introduced at the center of the system, the eigenvalues of the wavefunction under the \(\hat{C}_4\) rotation operation are modified. Taking a four-atom model with \(C_4\) symmetry as an example, after introducing a flux \(\Phi\), the operation on the eigenfunction gives
$
\hat{C}_4^{\Phi}|\psi\rangle=e^{i\frac{\Phi}{4}}e^{i\frac{2\pi n}{4}}|\psi\rangle.
$
Here, for \(\hat{C}_4(\Phi)\), we have
\begin{equation}
e^{i\theta/4}
\begin{bmatrix}
0 & 0 & 0 & 1 \\
1 & 0 & 0 & 0 \\
0 & 1 & 0 & 0 \\
0 & 0 & 1 & 0
\end{bmatrix}
\overset{\text{gauge equiv}}{\longrightarrow} \begin{bmatrix}
0 & 0 & 0 & e^{i\theta} \\
1 & 0 & 0 & 0 \\
0 & 1 & 0 & 0 \\
0 & 0 & 1 & 0
\end{bmatrix},
\end{equation}
where the left-hand expression represents the case where the total phase \(e^{i\theta}\) is uniformly distributed over all basis directions, while the right-hand expression concentrates the entire phase on a single basis direction. These two approaches are gauge-equivalent. In practical implementations, a more reasonable and experimentally feasible method is to distribute the gauge flux uniformly over the coupling phases along all four edges, thereby avoiding discontinuities caused by localized phase jumps. Furthermore, the introduced flux not only alters the eigenvalues under the \(\hat{C}_4\) rotation of the wave function but also changes the orbital angular momentum \(l_z\) if the wave function is an eigenstate of \(\hat{L}_z\). In this case, the orbital angular momentum after flux insertion needs to be revised as
$
l_z^{n}=\frac{4\left(\frac{\Phi}{4}+\frac{2\pi n}{4}\right)}{2\pi}.
$
With the modified eigenvalues \(E^{\prime}\) given by \(e^{\frac{\Phi}{4}},~ie^{\frac{\Phi}{4}},~-e^{\frac{\Phi}{4}},~-ie^{\frac{\Phi}{4}}\), the orbital angular momenta \(l_z^{n}\) become \(\frac{\Phi}{2\pi}+2,\; \frac{\Phi}{2\pi}+1,\; \frac{\Phi}{2\pi}+2,\; \frac{\Phi}{2\pi}-1\), and the corresponding orbital types of the wave functions are referred to as pseudo-\(s,\; p_+(p_x+ip_y),\; d,\; p_-(p_x-ip_y)\). When the artificial gauge flux \(\Phi\) is varied from 0 to \(2\pi\), not only do the symmetry eigenvalues undergo cyclic evolution, but the orbital angular momentum also evolves continuously, i.e., \(l_z^{0}\rightarrow l_z^{1}\rightarrow l_z^{2}\rightarrow l_z^{3}\rightarrow l_z^{0}\).

\begin{figure}[]
	\centering
	\includegraphics[width=0.5\textwidth]{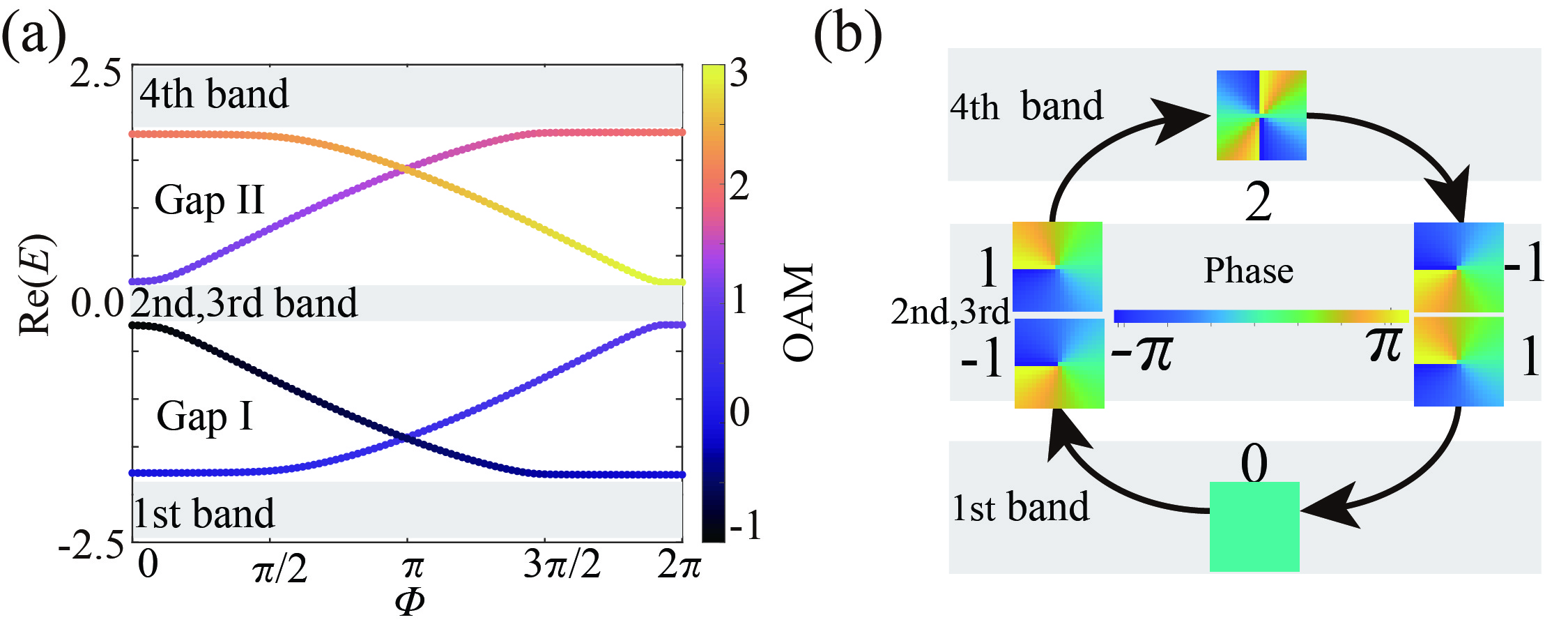}
\caption{(a) The band  diagram of the no gain and loss cavity. The topological cavity has gaps I and II, both of which contain topologically protected localized states crossing the gap. The localized states carry orbital angular momentum information of light and vary with the gauge flux $\Phi$. (b) The evolution of eigenstates along the cyclic spectral flows traversing gaps I and II.}
\label{fig:four_band}
\end{figure}

More importantly, as shown in Fig.~\ref{fig:four_band}(a), a pair of topological defect states appears in each of the band gaps I and II. This phenomenon can be verified by the presence of anomalous spectral filling and the calculation of a real-space topological invariant, which together reveal its nature as a topological Wannier cycle~\cite{Lin2022}. As shown in Fig.~\ref{fig:four_band}(b), when the gauge flux is continuously varied from \(0\) to \(2\pi\), clear spectral flow emerges in both band gaps, as evidenced by the evolution of the orbital angular momentum and phase from the initial to the final states. During this process, under time-reversal symmetry, the \(s\)-like state in the first band (with \(l_z = 0\)) evolves into the \(p_+\) state (with \(l_z = +1\)) in the second and third bands, while the \(p_-\) state (with \(l_z = -1\)) in the second and third bands evolves back into the \(s\) state. Similar cyclic evolution occurs for states in other bands as well, highlighting the continuous transformation of orbital angular momentum during the topological cycle.

\subsection{Complex eigenspectrum and  single-mode lasing}
\label{sec.Complex eigenspectrum1}
 
 Based on the two-dimensional SSH lattice model with an artificial gauge flux at the center, while also considering the non-Hermitian effect and nonlinear saturation mechanism arising from the pumped resonators (blue sites) and lossy resonators (red sites), as shown in Fig. \ref{fig:center_flux}, it is necessary to discuss first whether the introduced artificial gauge flux coupling will bring extra complexity to the formation and steady-state distribution of the nonlinear laser modes, thereby breaking the simplified analysis of the system dynamics. We analyze the local dynamic equations. The central gauge flux encircles four photonic fields in the lattice in a counterclockwise direction, with the corresponding lattice sites labeled sequentially as \(1,\,2,\,3,\,4\). Taking the photonic field at lattice site 1 as an example, its time evolution follows
 \begin{align}
i\dot{a}_{1}(t)& =J_2e^{i\frac{\mathit{\Phi}}{4}}a_2+J_2e^{-i\frac{\mathit{\Phi}}{4}}a_4\notag\\
&+(\frac{igP_{1}}{1+|a_{1}|^{2}/{I_{\text{sat}}}}-i\gamma)a_{1}+J_0a_{\text{outer}},\notag\\
-i\dot{a}_{1}^*(t) &=J_2e^{-i\frac{\mathit{\Phi}}{4}}a_2^*+J_2e^{i\frac{\mathit{\Phi}}{4}}a_4^*\notag\\
&+(\frac{igP_{1}}{1+|a_{1}|^{2}/{I_{\text{sat}}}}+i\gamma)a_{1}^*+J_0a_{\text{outer}}^*,
 \label{flux1}
\end{align}
where \(a_{\text{outer}}\) denotes all external photonic fields and \(J_0\) is the coupling strength between the photonic field at lattice site 1 and the external fields. Then, using $
i|\dot{a}_1(t)|^2 = a_1(t)^*\,i\dot{a}_{1}(t) + a_1(t)\,i\dot{a}_{1}^*(t),
$
we obtain
\begin{align}
i|\dot{a}_1(t)|^2&=J_2(e^{i\frac{\mathit{\Phi}}{4}}a_1^*a_2-e^{-i\frac{\mathit{\Phi}}{4}}a_1a_2^*)
\notag\\
&+J_2(e^{-i\frac{\mathit{\Phi}}{4}}a_1^*a_4-e^{i\frac{\mathit{\Phi}}{4}}a_4a_2^*)\notag\\
&+(\frac{2igP_{1}}{1+|a_{1}|^{2}/{I_{\text{sat}}}}-2i\gamma)|a_{1}|^2\notag\\
&+J_0(a_1^*a_{\text{outer}}-a_{\text{outer}}^*a_1).
 \label{complex_equ2}
 \end{align}
Extending this process to the remaining lattice photonic fields (2, 3, 4) and summing up the contributions from all four photonic fields for the Hermitian coupling term,
$\sum_{i=1}^{4}|\dot{a}_i(t)|^2 = 2J_2\,\mathrm{Im}\Big(e^{i\frac{\mathit{\Phi}}{4}}a_1^*a_2\Big) + 2J_2\,\mathrm{Im}\Big(e^{-i\frac{\mathit{\Phi}}{4}}a_2^*a_1\Big)+\dots,$
we find that although the phase factors \(e^{\pm i\frac{\Phi}{4}}\) contained in the flux coupling terms alter the instantaneous distribution of the photonic field, they ensure the conservation of the total probability of the entire system. This is manifested by the fact that, after summation, all the flux-related phase factors do not change the overall probability density but only modulate the distribution and flow of photons among the sites.

For the non-Hermitian coupling term,
\begin{align}
\sum_{i=1}^4|\dot {a}_i(t)|^2& = \sum_{i=1}^4 \left[ |a_{i}|^{2}\Big(\frac{gP_{i}}{1+|a_{i}|^{2}/I_{\text{sat}}}-\gamma\Big)\right]\notag\\
&+\sum_{i=1}^4 \left[ 2J_0\,\mathrm{Im}\Big(a_i^*a_{\text{outer}}\Big)\right],
\end{align}
it is evident that in this term the steady state of the laser mode is determined by the balance among the loss \(\gamma\), the gain \(gP_{i}\), and the nonlinear saturation \(I_{\text{sat}}\). Although the flux introduces local phase modifications and possible localization effects, it does not alter this core balance. In other words, the laser threshold, mode competition, and saturated output power are still governed by the gain-loss mechanism.

Furthermore, extending the analysis to the dynamic evolution of the system, we consider the evolved state \(\alpha_{m,n}^k(t) = a_{m,n}^k U_k(t),\)
where \(U_k(t)\) represents the time factor of the mode evolution. Its eigenvalue's imaginary part is given by
\begin{align}
\hbar \omega_{k,\text{imag}} = \sum_{m,n} |a_{m,n}^k|^{2}\left(\frac{gP_{m,n}}{1+|a_{m,n}^k|^{2}|U_k(t)|^{2}/I_{\text{sat}}}-\gamma\right),
\end{align}
where \(\omega_{k,\text{imag}}\) represents the growth rate of the \(k\)th eigenstate. It can be seen that under steady-state conditions, the magnitude of the eigenvalue's imaginary part reflects the gain-loss balance mechanism of the system, meaning that the stability of the final laser mode is determined by this balance. This also implies that the introduction of an artificial gauge flux, while adding local geometric phase and topological defect states, does not disrupt the global conservation of probability and the gain-loss balance. It also allows one to predict which modes will reach lasing threshold first and remain stable under the nonlinear saturation mechanism, thereby achieving steady-state laser output.

\begin{figure}[]
	\centering
	\includegraphics[width=0.5\textwidth]{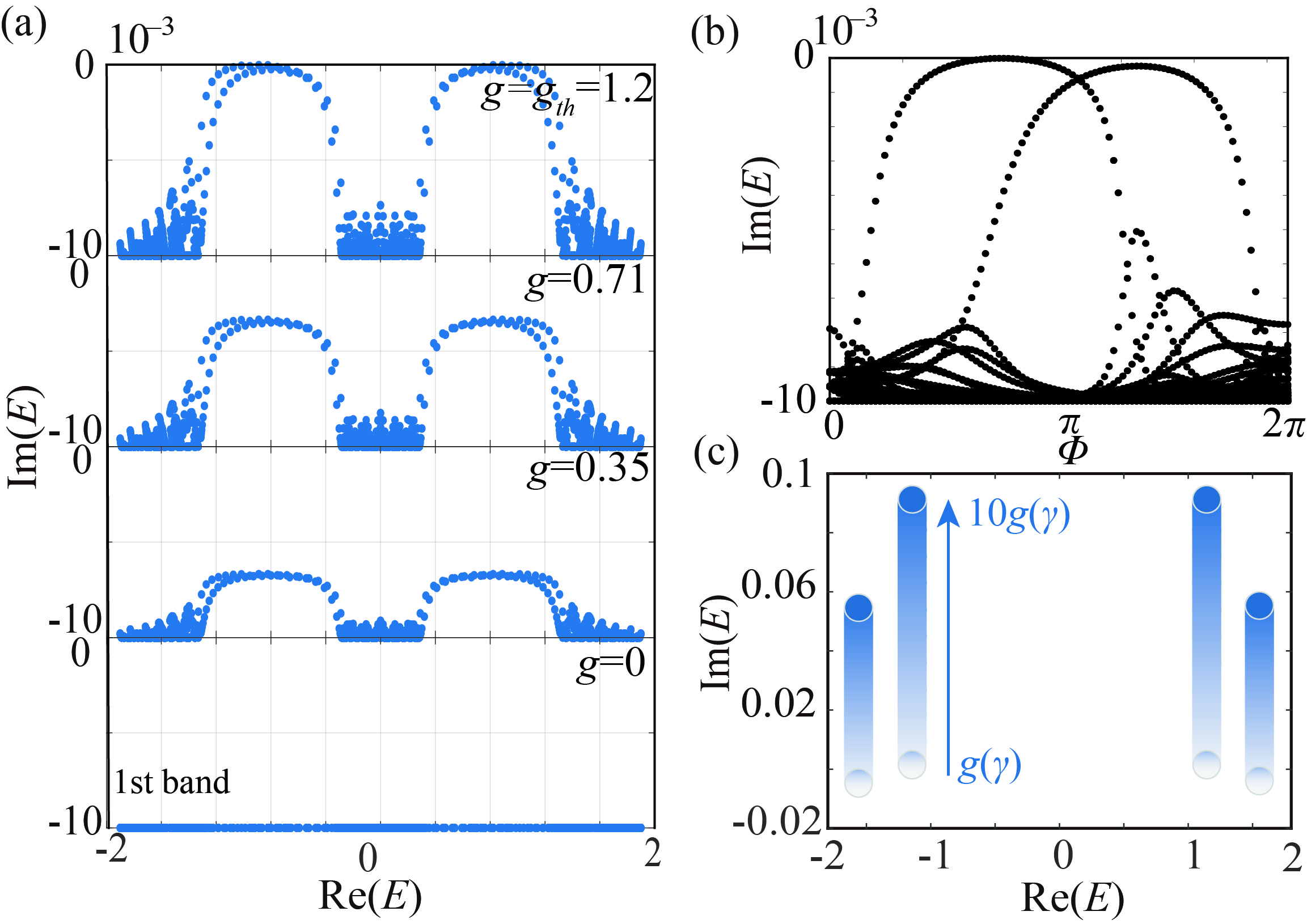}
	\caption{(a) Dependence of complex eigenvalues and gain-value. The imaginary eigenvalues of the four states evolve from zero gain to the threshold gain, are always the largest among all eigenstates, and become positive first at the threshold gain value $g_{\text{th}}=1.2$. (b) Relationship between imaginary energy and flux for threshold gain value  $g_{\text{th}} = 1.2$. The imaginary energy value is largest when the flux $\Phi=2.3876$. (c) Evolution of the complex eigenvalues of the topological laser modes as  the threshold gain increases at the flux $\Phi=2.3876$. The topological laser maintains single-mode lasing up to 10 times the threshold gain. We set $\gamma=0.01$ as the loss in the present work.}
\label{fig:real_imag1}
\end{figure}

Next, we analyze the complex energy spectrum of this model and the emergence of single-mode lasing. As shown in Fig. \ref{fig:real_imag1}(a), the complex eigenvalues are plotted as a function of the gain. Within band gaps I and II, as the gain increases from zero to the threshold, the imaginary parts of the eigenvalues for the localized states remain higher than those for the bulk states. When the gain reaches the threshold \(g_{\text{th}} = 1.2\), the imaginary parts of the eigenvalues of the two localized states become positive first. In addition, Fig. \ref{fig:real_imag1}(b) provides a more detailed depiction of the relationship between the imaginary eigenvalues and the magnetic flux, showing that the imaginary eigenvalues reach a maximum when the flux is \(\Phi = 2.3876\). Subsequently, when the gain significantly exceeds the threshold, the four modes with relatively large imaginary eigenvalues remain in single-mode lasing (i.e., as topological localized states), as shown in Fig. \ref{fig:real_imag1}(c). This is because the topological laser modes at the center are uniformly distributed, thereby depleting the available gain in the system and suppressing lasing in other cavity modes. This phenomenon is consistent with the conclusions in Refs. \cite{topological_laser1, PhysRevB.103.245305}.

\subsection{Mode carries OAM information and selection}
\label{subsec.Mode selection1}
The interesting characteristics exhibited by topological laser modes during their dynamic time evolution can be analyzed as follows. First, we define the initial state \(|\alpha_{m,n}^k(0)\rangle\). In practical applications, the eigenfunctions \(|a_{m,n}^k\rangle\) obtained under Hermitian conditions cannot be directly used as the laser initial state. Therefore, following the approach for selecting the initial state in Ref. \cite{PhysRevB.103.245305}, a global orbital angular momentum is designed such that the initial state is competitive, easily fabricated in experiments, and can selectively excite topological defect states. Specifically, we consider the projection weight of the initial state onto each eigenfunction,
$
|c_{m,n}^{k}|^2=|\langle a_{m,n}^k|\alpha_{m,n}^k(0)\rangle|^2,
$
where \(|c_{m,n}^{k}|^2\) represents the weight of the initial state on the \(k\)th eigenstate. By comparing with the eigenvalues of the eigenfunctions, one can directly observe the distribution of the initial state among the various modes in the system. If the projection weights of the initial state are significantly unbalanced, then the mode that is initially dominant will maintain its dominance during the evolution.

Next, we introduce the time-dependent evolution and define the fidelity to measure the overlap between the evolved state \( |\alpha_{m,n}^k(t)\rangle \) and the target eigenstate \( |a_{m,n}^k\rangle \)
\begin{align}
|C_{m,n}(t)|^2=\langle a_{m,n}^k|\alpha_{m,n}^k(t)\rangle,
\end{align}
where the wave function \( |\alpha_{m,n}^k(t)\rangle \) is normalized to ensure that the fidelity accurately reflects the overlap between the two states. More importantly, when \( |C_{m,n}(t)|^2 = 1 \), it indicates that in the steady state the orbital angular momentum information carried by the eigenstate \( |a_{m,n}^k\rangle \) is fully preserved, and that state represents the laser mode; when \( |C_{m,n}(t)|^2 = 0 \), it indicates that the evolved state is orthogonal to the target eigenstate, meaning that the information in that mode has been lost. Thus, calculating the fidelity allows us to assess the stability of each mode during the time evolution and reveals information about the final laser mode.

Before this, it is necessary to emphasize that the orbital angular momentum information carried by the eigenfunctions initially can be tracked to determine the orbital angular momentum information in the final output laser mode. Naturally, by applying the \(C_4\) rotation operation to the eigenfunctions,
$
\langle a_{m,n}^k|\hat{C}_4|a_{m,n}^k\rangle,
$
and calculating the resulting eigenvalues, one can identify the orbital angular momentum carried by all the eigenfunctions. In particular, we focus on the four topological defect states, whose eigenvalues are \(1, i, -1, -i\), corresponding to the orbital angular momenta \(l_z^{n} = n = 0,1,2,3(-1) \ (\text{mod} \ 4)\) and the orbital types of the eigenfunctions are \(s,\; p_+(p_x+ip_y),\; d,\; p_-(p_x-ip_y)\), respectively. The fidelities of these states during the evolution are denoted as \(|C_s|^2,\; |C_{p+}|^2,\; |C_d|^2,\; |C_{p-}|^2\).

Moreover, during the evolution, the artificial gauge flux at the center will redefine the orbital angular momentum of the wave function. The orbital angular momentum after the flux is inserted needs to be corrected as
$
l_z^{n}=\frac{4\left(\frac{\Phi}{4}+\frac{2\pi n}{4}\right)}{2\pi}.
$
With the eigenvalues after flux insertion \(E^{\prime}\) being \(e^{\frac{\Phi}{4}},\; ie^{\frac{\Phi}{4}},\; -e^{\frac{\Phi}{4}},\; -ie^{\frac{\Phi}{4}}\), the orbital angular momenta \(l_z^{n}\) become \(\frac{\Phi}{2\pi}+2,\; \frac{\Phi}{2\pi}+1,\; \frac{\Phi}{2\pi}+2,\; \frac{\Phi}{2\pi}-1\), and the corresponding orbital types of the wave functions are referred to as pseudo-\(s,\; p_+(p_x+ip_y),\; d,\; p_-(p_x-ip_y)\). Thus, we introduce the average orbital angular momentum of the different eigenstates during time evolution, which depends on the artificial gauge flux \(\Phi\),
\begin{equation}
\overline{\text{OAM}}(t)=\frac{\Phi}{2\pi}+2|c_d(t)|^2+|c_{p+}(t)|^2-|c_{p-}(t)|^2,
\label{eq:oam}
\end{equation}
which shows how the average orbital angular momentum evolves with time under a given initial state. It also reveals the stability of the average orbital angular momentum under the influence of the magnetic flux.

\begin{figure}[]
	\centering
	\includegraphics[width=0.45\textwidth]{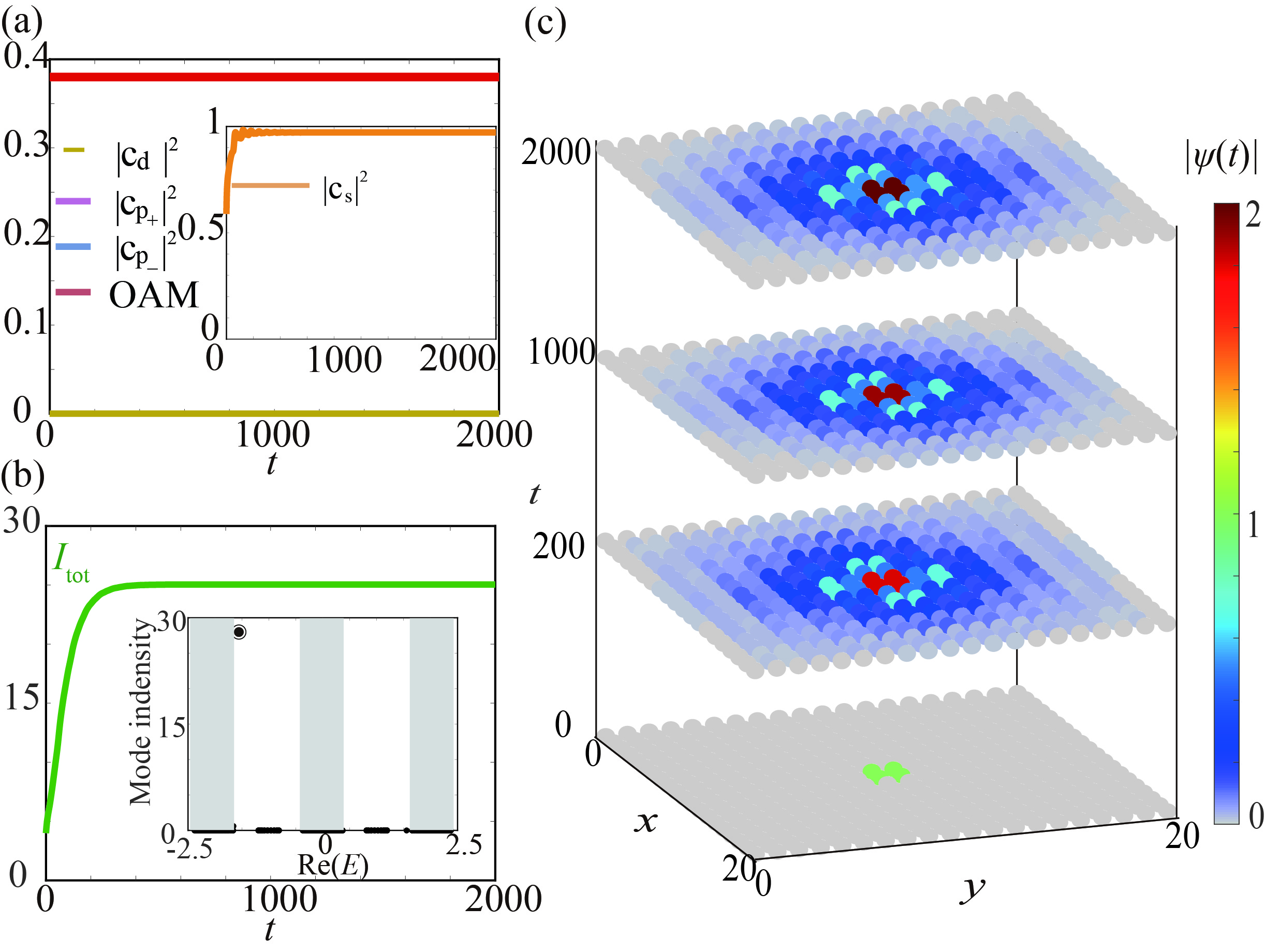}
	\caption{The dynamic evolution of the SSH lattice model with LGF configuration. (a) Time evolution of the fidelity of orbital angular momentum information carried by four localized states. The average OAM is represented by the solid red line. (b) Time evolution of the total intensity \(I_\text{tot}=\sum_{m,n}|\alpha_{m,n}^k(t)|^2\) . The inset is the final output laser mode. (c) Spatial distribution from initial state to final state. We set $\mathit{\Phi}=2.3876$ as the flux in the present work.}
\label{fig:time1}
\end{figure}

As shown in Fig. \ref{fig:time1}(a), it is observed that the fidelity depends solely on the choice of the initial state, with certain eigenstates persisting during the system evolution while others decay rapidly. The spatial distribution of the initial state is shown at \(t=0\) in Fig. \ref{fig:time1}(c), where small finite wave functions are assigned at the four central sites of the rectangular ring cavity, and the wave functions at other positions are zero. As seen in the inset of Fig. \ref{fig:time1}(a), the final output laser mode is the \(s\)-state with a fidelity \(|C_s(t)|^2 = 1\), while the fidelities of the other three modes quickly drop to 0. Interestingly, under this initial state, the average orbital angular momentum is solely influenced by the gauge flux and remains stable throughout the evolution, as indicated by the red solid line.

Furthermore, Fig. \ref{fig:time1}(b) shows the total intensity \(I_\text{tot}=\sum_{m,n}|\alpha_{m,n}^k(t)|^2\) at any given time. It can be seen that the system evolves to a steady state after a sufficiently long time. Finally, as shown in the inset of Fig. \ref{fig:time1}(b), the system eventually converges to an \(s\)-state laser mode. The amplitude distribution of the wave function in the rectangular ring cavity at different times is also provided. As shown in Fig. \ref{fig:time1}(c), at \(t=2000\) the system has reached a steady state, with the wave function amplitude distribution of the laser mode spreading from the four central sites with small amplitudes to very small amplitudes in the surrounding area.

\section{EXTENDED GAIN AND GAUGE FLUX (EGF)}
\label{sec:EXTENDED GAIN AND GAUGE FLUX (EGF)}
\subsection{Topological Wannier cycle in EGF}
\label{sec.Topological Wannier cycle in LGF}
\begin{figure}[]
	\centering
	\includegraphics[width=2.5in]{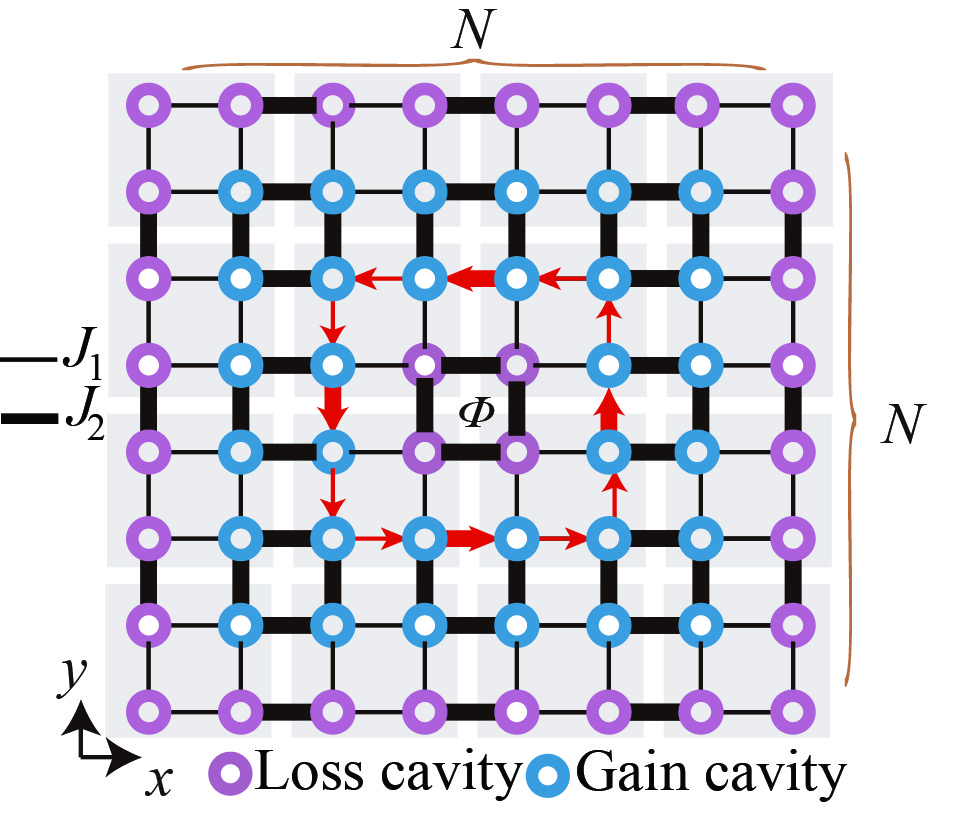}
\caption{A planar Lattice with extend gauge flux representation of the SSH model in Eq. (\ref{Hamiltonian_total}) with open boundaries in both directions. The sites has a particle loss (red) and a particle gain (blue). $\mathit{\Phi}$ is an adjustable periodic magnetic flux in the middle of the plate, the $J_1$ and $J_2$ is the intra-cell and inter-cell coupling, respectively.}
\label{fig:six_model}
\end{figure}

In addition, this work further investigates the relationship between the size of the region where the gauge flux is introduced and the number of topological Wannier cycles. This not only enhances experimental feasibility but also may lead to new topological phenomena. To this end, the range of the localized gauge flux is further expanded, and a new extended gauge flux is proposed in the 2D SSH lattice model, as shown in Fig. \ref{fig:six_model}. Furthermore, when the magnetic flux is varied, the symmetry eigenvalues of the wave function under the \(C_4\) operation will also change. In the original four-atom model that satisfies \(C_4\) symmetry, it becomes a \(4\times 4\) atomic model; when a magnetic flux \(\Phi\) is introduced, acting on the eigenfunctions still gives
$C_4^{(\Phi)}|\psi\rangle=e^{\frac{\Phi}{4}}e^{\frac{2\pi n}{4}}|\psi\rangle.$
Unlike the previous case of introducing a central gauge flux, in this case the phase accumulated over the entire loop is \(\Phi\), while on each edge the accumulated phase is \(\frac{\Phi}{6}\). Interestingly, the coupling edges that form the gauge flux are cyclically connected with coupling strengths varying from weak to strong and then back to weak.

\begin{figure}[]
	\centering
	\includegraphics[width=3in]{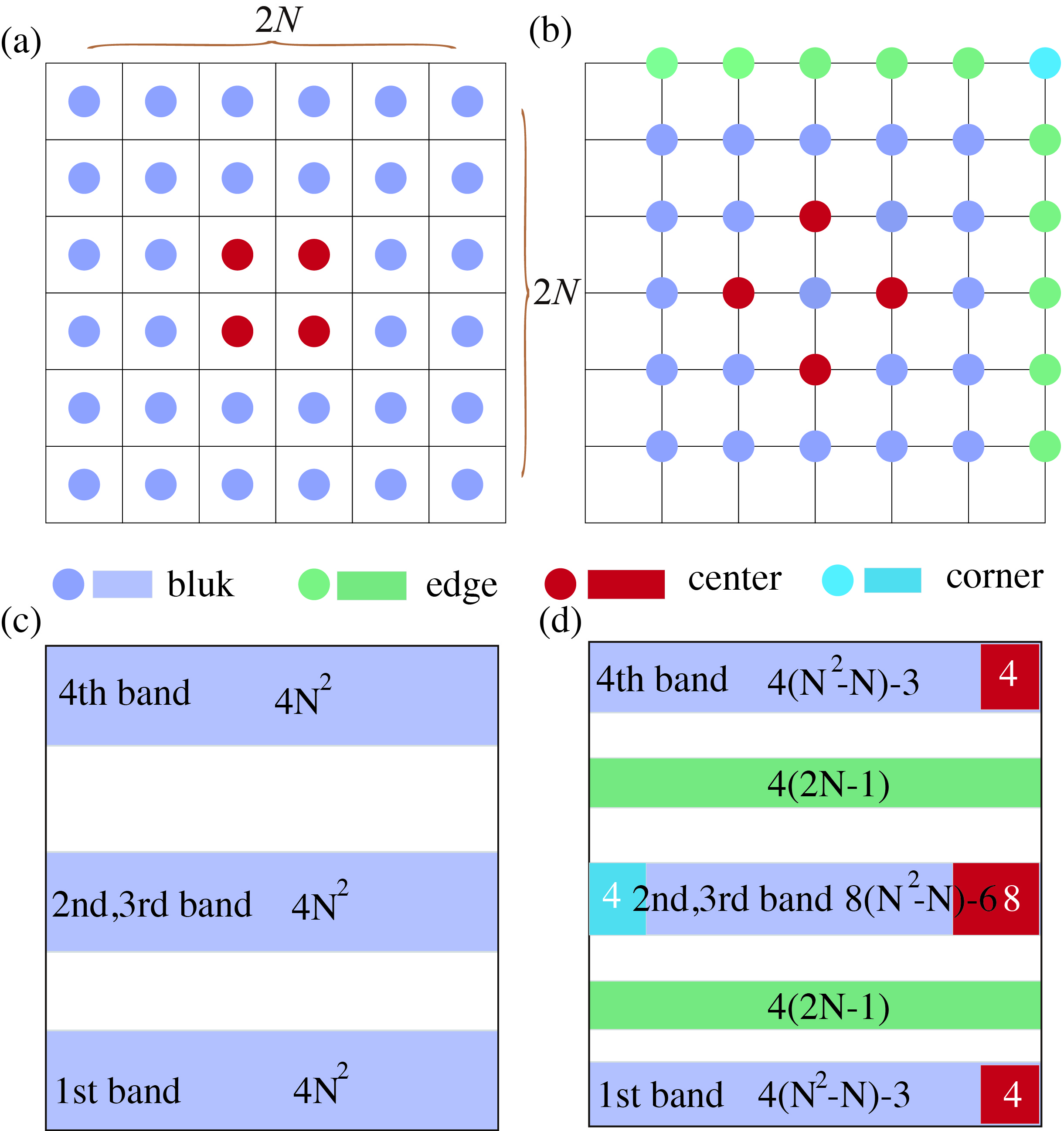}
\caption{Anomalous filling phenomenon in the 2D SSH model with extended gauge flux. (a) and (c) Schematic diagrams of the Wannier center distribution and supercell energy spectrum of the 2D SSH model in the trivial phase; (b) and (d) Schematic diagrams of the Wannier center distribution and supercell energy spectrum of the 2D SSH model in the topological phase. The numbers of bulk, edge, localized, and corner states are labeled in purple, green, red, and blue, respectively.}
\label{fig:tianchong_model}
\end{figure}

Topological Wannier cycles involve two important factors: one is the intrinsic factor, namely, the anomalous filling. Specifically, Wannier orbitals occupying non-unit-cell centers can directly reflect the number of bulk states and the asymmetry of the lattice. The other factor is the decisive one the real-space topological number which essentially characterizes the nontrivial distribution of Wannier orbitals in real space. First, in the topologically equivalent limit where intracell weak coupling vanishes ($J_1\rightarrow 0$), we investigate how increasing the spatial scale of the gauge flux in the 2D SSH model modifies the distribution of Wannier orbitals centered outside unit cells, as illustrated in Fig. \ref{fig:tianchong_model}. Crucially, these findings maintain adiabatic equivalence even for finite intracell coupling ($J_1\neq 0$). In Fig. \ref{fig:tianchong_model}(a), when the 2D SSH model is in the trivial phase, all Wannier orbitals occupy the unit-cell centers, which implies the absence of filling anomaly and no gapless spectral flow. And this implies that there are no edge or corner states in the band gap, and the number of bulk states in the four Bloch bands are integer multiples of 4 due to symmetry, as shown in Fig. \ref{fig:tianchong_model}(c). However, when the 2D SSH model is in the topological phase, the Wannier orbitals are shifted toward the corners (see Fig. \ref{fig:tianchong_model}(b)). It is found that, in the limiting case, most bulk states have Wannier orbitals that can always form complete irreducible representations in groups of four under \(C_4\) symmetry, including the states marked in green and blue. In particular, the four Wannier orbitals at the center (or the four relatively isolated bulk states in the limit), marked in red, cannot form a complete \(C_4\) irreducible representation individually. This implies that, under the supercell configuration, each band has an extra isolated bulk state. In the  topological phase, the spatial distribution of these four Wannier centers is coupled with the configuration of EGF in the lattice, with this coupling primarily manifested in the inter-cell strong coupling strength. However, the appearance of the four Wannier centers means that there will be four fractional corner charges of \(1/4\). The numbers of edge states and corner states distributed in the band are shown in Fig. \ref{fig:tianchong_model}(d). It is worth mentioning that when probing the properties of the band structure in actual optical or acoustic systems, the edge states or corner states may tend to merge into the bulk states.

After analyzing the anomalous filling phenomenon, we next discuss how the real-space topological number reveals the number of topological Wannier cycles. Recall that when a central gauge flux is introduced into the 2D SSH model in the topological phase, one topological Wannier cycle occurs. The real-space topological number can still be defined as~\cite{Lin2022,doi:10.1126/science.aaz7650}
\begin{align}
&(\delta_1,\delta_2,\delta_3,)\notag\\
&=(m(E_{-1})-m(E_0),m(E_{1})-m(E_0),m(E_{1})-m(E_0)),
\end{align}
where \( m(E_n) \) denotes the number of supercell eigenstates with the rotation eigenvalue \(E_n\) in the target band gap. If this real-space topological number is nonzero, then there is an anomalous filling phenomenon in the energy spectrum.

It should be noted that when the extended gauge flux (4×4) is introduced, each rotation eigenvalue (such as \(E_0\), \(E_1\), \(E_2\), \(E_{-1}\)) is four-fold degenerate. If, for example, the number of supercell eigenstates with the \(E_0\) rotation eigenvalue in the target band gap is 4, and similarly for \(n=1,2,3\), this implies that four topological Wannier cycles occur simultaneously. For instance, for the first band gap, if the gauge flux is varied from 0 to \(2\pi\), Ref.~\cite{Lin2022} gives \(\Delta(E_0)=-1\), \(\Delta(E_1)=1\), \(\Delta(E_2)=0\), \(\Delta(E_{-1})=0\). This means that below the band gap there is always one extra eigenstate with the \(E_0\) eigenvalue, and above the band gap there is one extra eigenstate with the \(E_1\), \(E_{-1}\), and \(E_2\) eigenvalues, respectively. However, in our system, since each rotation eigenvalue (such as \(E_0\), \(E_1\), \(E_2\), \(E_{-1}\)) is four-fold degenerate, this means that below the first band gap there are four extra eigenstates with the \(E_0\) eigenvalue, and above the band gap there are four extra eigenstates with each of the \(E_1\), \(E_{-1}\), and \(E_2\) eigenvalues. The analysis for the second band gap is similar; hence, by combining the real-space topological numbers of the two band gaps, one can deduce that the first band has four extra \(E_0\) states, the second and third bands have four pairs of \(E_{\pm1}\) states, and the fourth band has four extra \(E_2\) states, which is consistent with the previously analyzed anomalous filling phenomenon. Therefore, the spatial scale of the introduced gauge flux influences the number of topological Wannier cycles.

\begin{figure}[]
	\centering
	\includegraphics[width=3.5in]{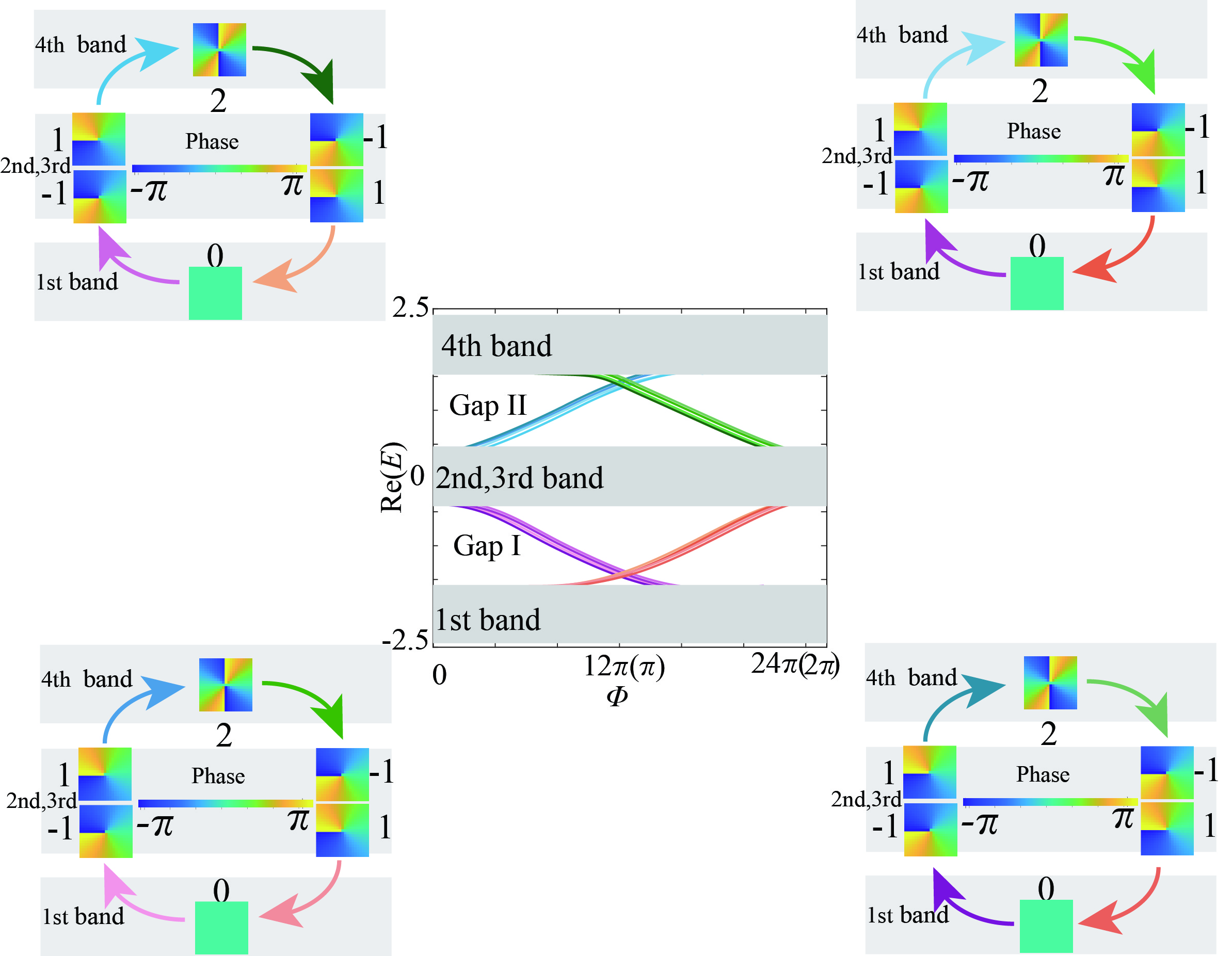}
\caption{Topological Wannier cycles in the 2D SSH model with extended gauge flux. Middle: the supercell energy spectrum evolves with the gauge flux varying from 0 to \(2\pi\), exhibiting four spectral flow events in the band gaps. Four corner panels: schematic diagrams of the evolution of different eigenstates and orbital angular momenta.}
\label{fig:wanier_four}
\end{figure}

Moreover, after introducing the extended gauge flux into the 2D SSH model, the evolution of the supercell band gaps with respect to the magnetic flux is shown in the central panel of Fig. \ref{fig:wanier_four}. In band gaps I and II, four pairs of topological defect states emerge, respectively. Meanwhile, the four corner panels of Fig. \ref{fig:wanier_four} display four distinct spectral flow events, indicating the presence of multiple topological Wannier cycles. Similarly, it is found that as the gauge flux varies from 0 to \(2\pi\), by comparing the orbital angular momentum and phase of the initial and final states, four spectral flow events occur in the two band gaps. Specifically, during the variation of the gauge flux from 0 to \(2\pi\), considering time-reversal symmetry, the four \(s\)-states (with \(l_z=0\)) from the first band simultaneously evolve into four \(p_+\)-states (with \(l_z=+1\)) in the second and third bands, and the four \(p_-\)-states (with \(l_z=-1\)) in the second and third bands simultaneously evolve back into four \(s\)-states (with \(l_z=0\)); similar discussions apply to states in other bands. This number of topological Wannier cycles is closely related to the spatial scale of the introduced gauge flux. For this reason, a quantitative relation is also provided: the gauge flux contains four Wannier cores, which means that there are four topological Wannier cycles.

\subsection{Complex eigenspectrum and  single-mode lasing}

\begin{figure}[]
	\centering
	\includegraphics[width=3.45in]{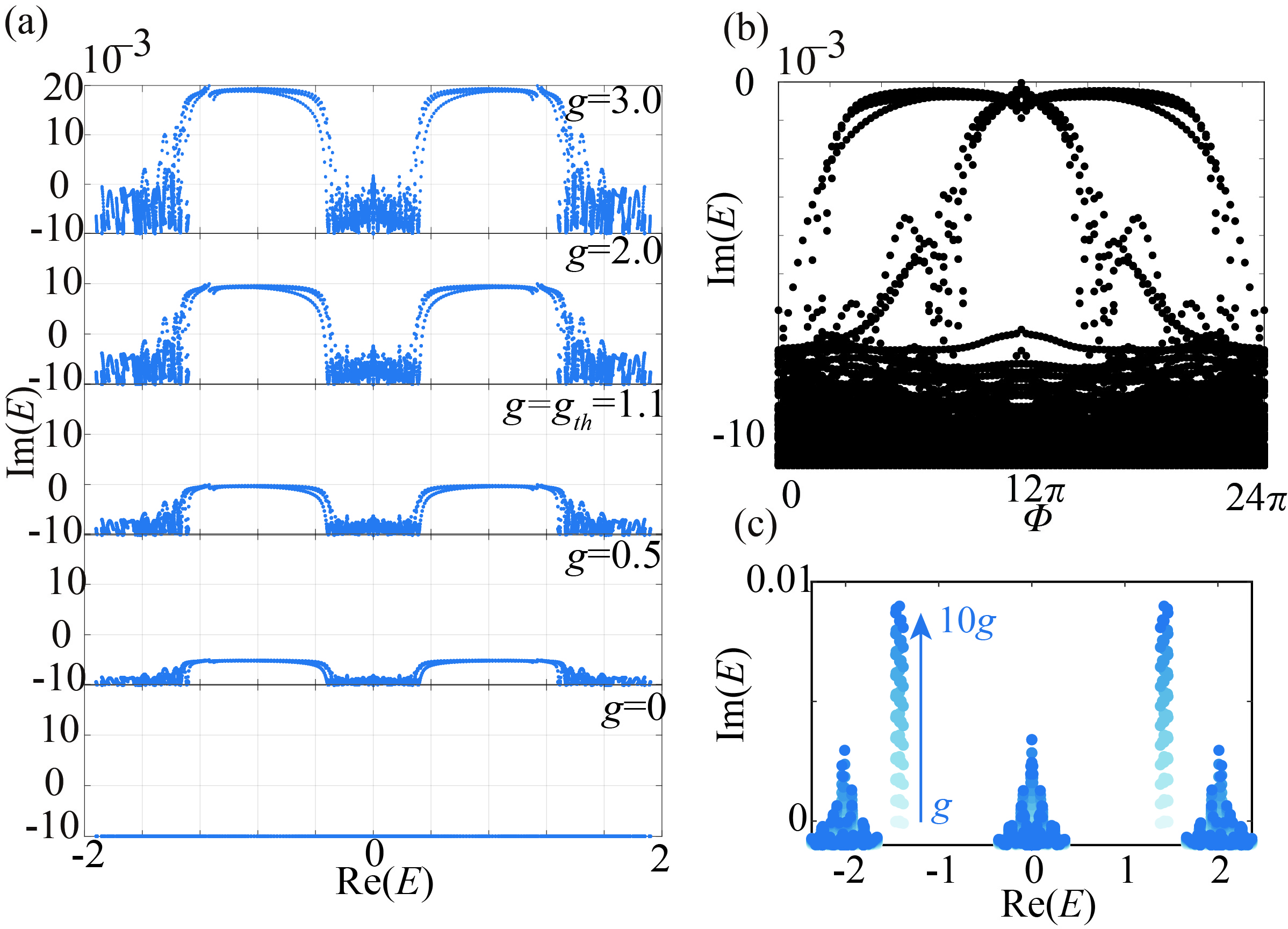}
	\caption{(a) Dependence of complex eigenvalues and gain-value. The imaginary eigenvalues of the four states evolve from zero gain to the threshold gain, are always the largest among all eigenstates, and become positive first at the threshold gain value $g_{\text{th}}=1.1$. (b) Relationship between imaginary energy and flux for threshold gain value gth = 1.1. The imaginary energy value is largest when the flux $\Phi = 12\pi \pm \Delta \Phi$. (c) Evolution of the complex eigenvalues of the topological laser modes as  the threshold gain increases at the flux $\Phi = 12\pi \pm \Delta \Phi$. The topological laser maintains single-mode lasing up to 10 times the threshold gain. We set $\gamma=0.01$ as the loss in the present work. }
\label{fig:tuozhanfou}
\end{figure}

Similarly, we further analyze the complex energy spectrum and single-mode lasing of the 2D SSH lattice with the extended gauge flux. Consider the complex energy spectrum of the 2D SSH lattice model with the new extended artificial gauge field \(\Phi\) in the linear limit of Eq. (\ref{laser_dynamics}) (i.e., \(I_{\text{sat}} \rightarrow 0\)). Figure \ref{fig:tuozhanfou}(a) shows the variation of the complex energy spectrum as a function of the gauge flux \(\Phi\) for different gain values \(g\) (ranging from 0 to 3), here considering the total gauge flux as \(\Phi=24\pi\). The results indicate that when the gain reaches the threshold \(g_{th} = 1.1\), the topological defect states in band gaps I and II begin to acquire positive imaginary parts of their eigenvalues, meaning that the topological defect states are the first to reach lasing oscillation. Furthermore, above the threshold, these localized states remain within the corresponding band gaps, with growth rates significantly faster than those of other modes, and their eigenvalue imaginary parts consistently remain higher than those of the other states.

Interestingly, we also examine the variation of the imaginary part of the eigenvalues with respect to the magnetic flux at the threshold (see Fig. \ref{fig:tuozhanfou}(b)), which shows that the imaginary part is symmetric about \(\Phi = 12\pi(\pi)\). This indicates that under variations of the gauge flux, the Hamiltonian still satisfies time-reversal symmetry and the system's dissipation has reached a balanced state. Therefore, it can be concluded that at \(\Phi = 12\pi(\pi)\), the imaginary parts of the eigenvalues of the laser modes reach their maximum values. However, this is the critical value of the eigenvalue's imaginary part and should not be used as the sole criterion for judgment. Instead, it is observed that at \(\Phi = 12\pi(\pi) \pm \Delta \Phi\) (where \(\Delta \Phi\) denotes the difference in magnetic flux), the imaginary parts of the eigenvalues of the laser modes each reach their maximum.
Figure \ref{fig:tuozhanfou}(c) further shows the complex energy spectrum when the gain is increased by a factor of 10 to the threshold, under the condition \(\Phi = 12\pi(\pi) \pm \Delta \Phi\). Under these conditions, the laser modes always maintain the largest eigenvalue imaginary parts compared to the other modes.

\subsection{Mode carries OAM information and selection}
\begin{figure}[]
	\centering
	\includegraphics[width=3.45in]{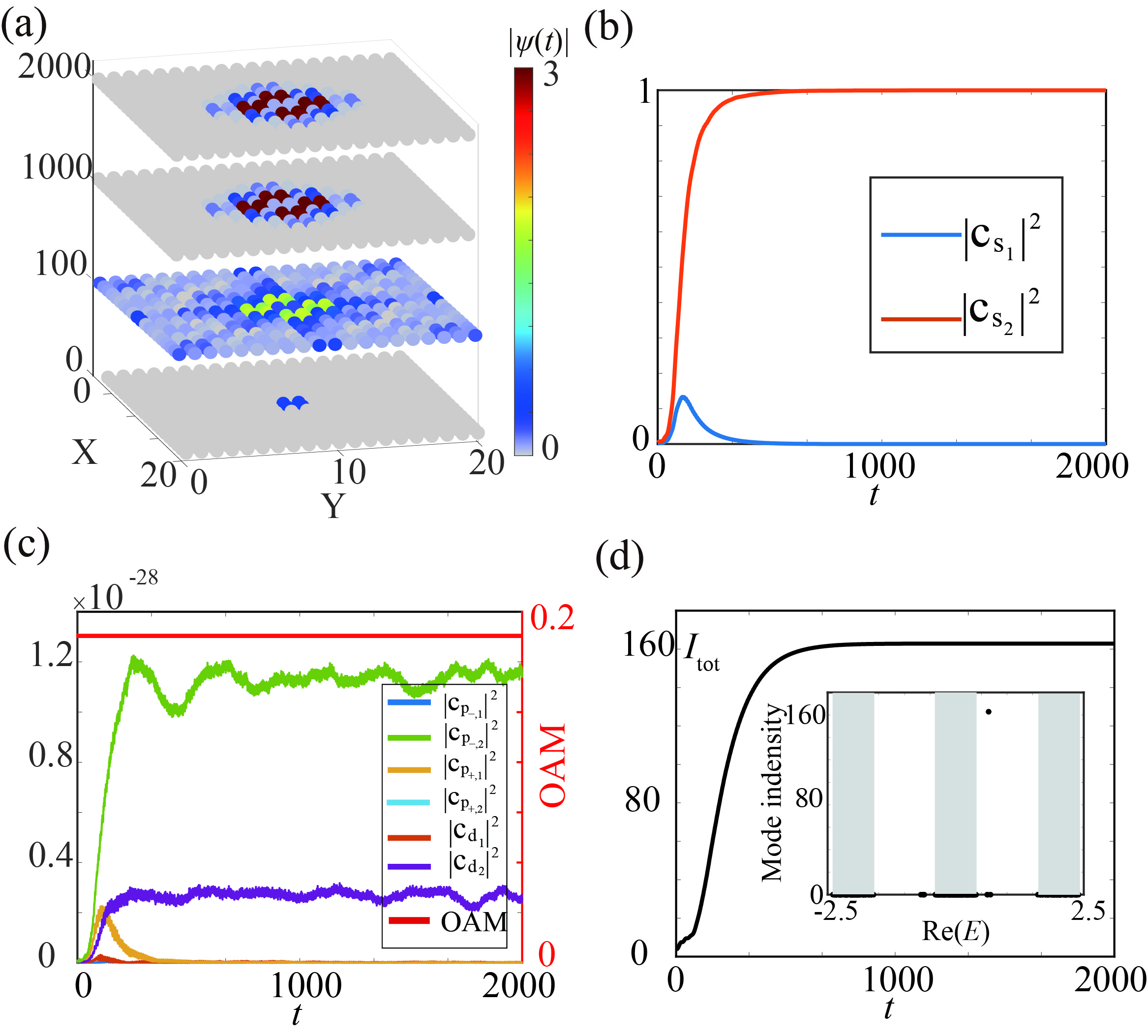}
	\caption{The dynamic evolution of the LGF configuration. (a) Time evolution of the fidelity of orbital angular momentum information carried by four localized states. The average OAM is represented by the solid red line. (b) Time evolution of the total intensity \(I_\text{tot}=\sum_{m,n}|\alpha_{m,n}^k(t)|^2\). The inset is the final output laser mode. (c) Spatial distribution from initial state to final state. We set $\Phi=2.3876\times12$ as the flux in the present work.}
\label{fig:time_oam22}
\end{figure}

The interesting characteristics exhibited by the topological laser modes during their dynamic time evolution can be analyzed as follows. To ensure that the laser mode carries complete orbital angular momentum (OAM) information, the choice of the initial state is crucial. By setting the gauge flux to \(\Phi= 2.3876 \times 12\), Figs. \ref{fig:time_oam22}(b) and (c) show the evolution of the fidelity of 8 localized states over time; the remaining 8 localized states are not considered because their energies are closer to those of the bulk states. The two \(s\)-states, \(\lvert a_{s},1\rangle\) and \(\lvert a_{s},2\rangle\), which carry the OAM information, compete with each other during the early stage of evolution. However, after some time, the fidelity of \(\lvert a_{s},1\rangle\) stabilizes and reaches 1, while that of \(\lvert a_{s},2\rangle\) decreases to 0. For the other 6 states, their fidelities also drop rapidly to 0, although they similarly show a competitive trend. Unlike the LGF model discussed earlier, which involved competition among only 4 states, the competition among localized states here is more complex due to the interactions and selectivity among multiple states, which helps in understanding the stability and selectivity of localized states in non-Hermitian and nonlinear systems.

In the 2D SSH model with the extended gauge flux, due to the change in the total gauge flux, the orbital angular momentum \(l_z^{n}\) needs to be redefined as 
$
\frac{\Phi}{12\times2\pi}+2,\; \frac{\Phi}{12\times2\pi}+1,\; \frac{\Phi}{12\times2\pi}+2,\; \frac{\Phi}{12\times2\pi}-1,
$
and the corresponding orbital types of the wave functions are classified as pseudo-\(s\), \(p_+(p_x+ip_y)\), \(d\), and \(p_-(p_x-ip_y)\), respectively. Moreover, since each \(C_4\) rotation eigenvalue now corresponds to a four-fold degenerate state, the original expression for the average orbital angular momentum (Eq. (\ref{eq:oam})) needs to be further modified to
\begin{equation}
\overline{\text{OAM}}(t)=\sum_{n=1}^2\Bigl(\frac{1}{12}\frac{\mathit{\Phi}}{2\pi} + 2\lvert c_{d,n}(t)\rvert^2 + \lvert c_{p+,n}(t)\rvert^2 - \lvert c_{p-,n}(t)\rvert^2\Bigr).
\end{equation}
As shown by the red solid line in Fig. \ref{fig:time_oam22}(c), this result is consistent with that of the previous LGF configuration. The average orbital angular momentum remains solely influenced by the gauge flux, with a steady-state value of approximately \(\overline{\text{OAM}}(t)=0.19\).

The final output laser mode is \(\lvert a_{s},1\rangle\) (as shown in the inset of Fig. \ref{fig:time_oam22}(d), which is consistent with its fidelity reaching 1; meanwhile, the total intensity \(I_{\text{tot}}\) of all states reaches 160. As the system evolves towards a steady state, the intensity saturates (see Fig. \ref{fig:time_oam22}(d)). Figure \ref{fig:time_oam22}(a) presents the evolution of the spatial field distribution over time: as the evolution proceeds, the field density becomes increasingly concentrated in the central region, forming a high-intensity core, while the surrounding regions remain relatively weak. This indicates that the system is transitioning towards a stable laser mode. Compared with the previous LDG model, the high-intensity region at the center in this case is significantly expanded, leading to a more concentrated laser output.

\section{Modulated OAM and mode intensity}
\label{subsec.Mode selection2}
Based on the proposed 2D SSH lattice model with artificial gauge fluxes introduced at two different spatial scales, it is demonstrated that the emerging defect states are topologically characterized, and that the number of topological Wannier cycles is related to the spatial scale of the gauge flux. More importantly, when the introduced artificial gauge flux varies from 0 to \(2\pi\), the orbital angular momentum of the photonic field in the system also undergoes a continuous cyclic evolution. However, if non-Hermitian effects and nonlinear saturation mechanisms are taken into account, can the laser still achieve continuously tunable orbital angular momentum? To address this, we consider the relationship between the laser mode and the gauge flux. Specifically, the average orbital angular momentum when the same initial state evolves in time to a steady state as the gauge flux varies from 0 to \(2\pi\).

\begin{figure}[]
	\centering
	\includegraphics[width=0.45\textwidth]{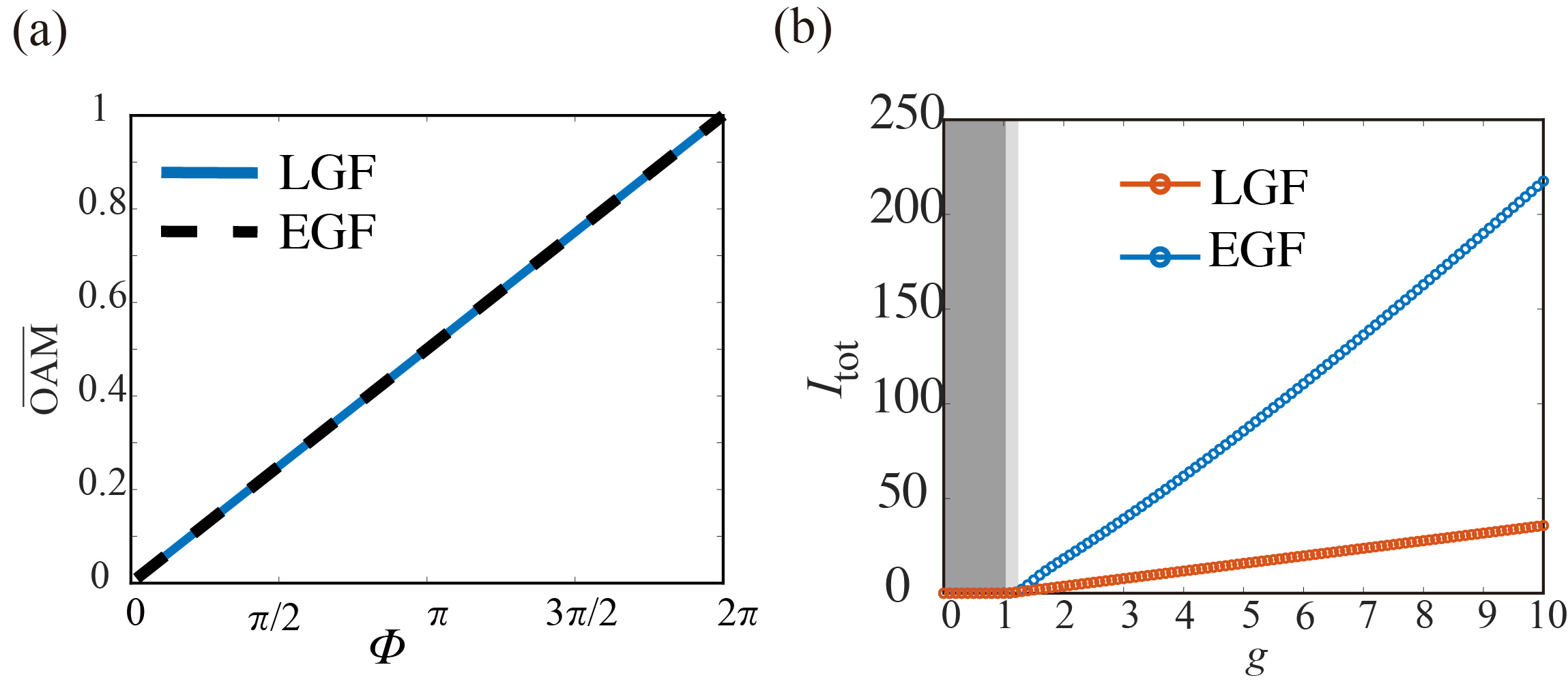}
	\caption{(a) Averaged relationship between OAM and flux in the output $s$-state. The blue solid line and the black dashed line represent LGF and EGF configurations, respectively. (b) Intensity-gain relation of the LGF and EGF configurations with the flux $\Phi=2.3876(2.3876\times12)$. The gray shaded area represents the area below the gain threshold. }
\label{fig:time_oam2}
\end{figure}

Based on the proposed LGF and EGF model structures, it is demonstrated that lasers with continuously tunable nonzero orbital angular momentum can be realized (as shown in Fig. \ref{fig:time_oam2}(a)). In both models, as the gauge flux varies from 0 to \(2\pi\), the average orbital angular momentum evolves smoothly from 0 to 1. This evolution is attributed to the fact that the final laser mode always remains in the \(s\) state, which is characterized by zero orbital angular momentum, so that the overall average orbital angular momentum is solely controlled by the gauge flux. Notably, the initial state can also be modified such that the average orbital angular momentum depends not only on the magnetic flux but also on the fidelity of the \(d\), \(p_+\), and \(p_-\) states.

Furthermore, to reflect the output power and efficiency of the laser under the two configurations (LGF and EGF), the total intensity at the steady state, \(I_{\text{tot}}\), is calculated (see Fig. \ref{fig:time_oam2}(b)). In the LGF model, when the gain is below the threshold \(g_{\text{th}} = 1.2\), the total intensity remains zero; once the gain exceeds the threshold, the total intensity increases slowly and linearly with the gain. For example, at \(g=10\gamma\), the total intensity reaches only \(I_{\text{tot}} = 35.73\), which is relatively low compared to the LGF model. In contrast, in the EGF model, when the gain reaches \(g=10\gamma\), the total intensity significantly increases to \(I_{\text{tot}} = 217.60\); moreover, even at a higher gain of \(g=50\gamma\), the output remains stable, with the total intensity rising to \(I_{\text{tot}} = 1200\). This significant increase in intensity highlights the outstanding performance of the EGF model under high-gain conditions, demonstrating that its laser output is markedly superior in efficiency and stability compared to the LGF model.

\section{single-mode area}
\label{sec:single-mode area}
The effective single-mode area is used to characterize the spatial region over which a beam propagates stably in optical waveguides, fibers, or laser cavities \cite{single_are1, single_are2, single_are3}. This physical quantity quantitatively describes the effective cross-sectional area of the optical mode, providing key information on the mode’s spatial confinement and energy distribution. Typically, a smaller single-mode area indicates stronger confinement and higher optical intensity, while a larger area suggests weaker confinement and lower intensity. Therefore, under steady-state conditions, the effective single-mode area is defined as \cite{single_are3}
\begin{equation}
A_{\text{eff}}=\frac{\left(\sum_{m,n}\left|a_{m,n}\right|^2\,\right)^2}{\sum_{m,n} \left|a_{m,n}\right|^4\,},
\end{equation}
where \(a_{m,n}\) represents the amplitude of the photonic field and \(|a_{m,n}|^2\) is the intensity of the laser mode. Subsequently, the effective single-mode areas of the laser modes for the LGF and EGF models were calculated to be 11.45 and 20.17, respectively. This indicates that the effective single-mode area of the EGF model is much larger than that of the LGF model, meaning that the excitation of the laser mode is more concentrated and efficient, which can enhance the overall performance of the laser system. It is noteworthy that, for simplicity, the number of unit cells at each boundary in the rectangular supercell is assumed to be uniform and is set to 1.

Additionally, the total area of the system is 400. By considering the ratio of the effective single-mode area to the total area, it was found that for the first mode (with an effective area of 11.45), it occupies approximately 2.87\% of the total area, whereas for the second mode (with an effective area of 20.17), this ratio increases to about 5.04\%. These values highlight the high efficiency of mode confinement, indicating that laser emission is achieved with only a small fraction of the total area. This further confirms the superiority of the design in terms of spatial efficiency and its ability to concentrate the output intensity within a confined region.

\section{DISCUSSION AND CONCLUSION}
\label{sec:DISCUSSION AND CONCLUSION}
In this work, we proposed and investigated a novel framework for topological lasers based on a two-dimensional SSH model, incorporating two configurations: LCG and ECG. By leveraging topological Wannier cycles, we demonstrated robust spectral flows traversing the band gaps, uncovering the fundamental connection between real-space topological invariants and eigenstate evolution. Our results emphasize the critical role of artificial gauge flux in dynamically controlling topological states, offering new possibilities for experimental realization in optical systems.

The key findings of this study include the tunability of OAM in both LGF and EGF configurations. Through fidelity analysis and spectral evolution, we showed that OAM information in laser modes can be effectively encoded and preserved, even under non-Hermitian and nonlinear conditions. Notably, the EGF configuration exhibited six times higher single-mode laser intensity compared to the LGF configuration, along with a larger single-mode area. These performance advantages highlight the potential of the EGF configuration for achieving efficient and stable laser outputs as well as precise OAM dynamic control.

Our design is primarily suitable for implementation on photonic platforms. In optical systems, LGF and EGF configurations can be realized using silicon-based photonic crystals integrated with gain and loss regions. The introduction of artificial gauge flux can be achieved by dynamically modulating the coupling constants of photonic crystal waveguides or by employing synthetic dimension methods. Experimentally, the spectral flows induced by topological Wannier cycles and the OAM dynamics can be directly detected by measuring the far-field diffraction patterns or using interferometric techniques. Moreover, the EGF configuration's reduced sensitivity to gain distribution makes it more feasible for experimental implementation. In particular, the remarkable OAM tunability exhibited by the EGF configuration provides a novel mechanism for generating vortex beams with precisely controlled angular momentum. Such vortex beams hold significant potential for applications in optical communications, quantum optics, and super-resolution imaging.

\begin{acknowledgments}
This work was supported by the National Key R\&D Program of China (2022YFA1404400), the National Natural Science Foundation of China (Grant No. 12125504), the ``Hundred Talents Program'' of the Chinese Academy of Sciences, the Priority Academic Program Development (PAPD) of Jiangsu Higher Education Institutions.
\end{acknowledgments}


\bibliography{NH_chiral_laser}

\end{document}